\documentclass[acmsmall,screen]{acmart}
\AtBeginDocument{%
  }
\usepackage{xspace}
\usepackage{tikz}
\usepackage{booktabs}
\usepackage{tabularx}
\usepackage{wrapfig}
\usepackage{todonotes}
\usepackage{enumitem}
\usepackage{booktabs}
\usepackage[table]{xcolor}
\usepackage{makecell}

\definecolor{inquiryred}{HTML}{A12B39}
\definecolor{requestblue}{HTML}{245A9C}
\definecolor{replygreen}{HTML}{28734A}
\usetikzlibrary{arrows.meta}
\usepackage{fontawesome5}
\usepackage[most]{tcolorbox}
\setcopyright{none}
\newtcbox{\methodtag}[1][black!8]{
  on line,
  colback=#1,
  colframe=#1,
  boxrule=0pt,
  arc=8pt,
  boxsep=0pt,
  left=6pt,
  right=6pt,
  top=2pt,
  bottom=2pt,
  fontupper=\sffamily\footnotesize\bfseries
}
\acmISBN{978-1-4503-XXXX-X/2018/06}

\newcommand{\eg}{\textit{e.g.,}\xspace}
\newcommand{\etal}{\textit{et al.}\xspace}

\newcommand{\smalltt}[1]{\texttt{\small{#1}}}

\newcommand{\supervisor}{\textit{\textcolor{requestblue}{Supervisor}}\xspace}
\newcommand{\supervisors}{\textit{\textcolor{requestblue}{Supervisors}}\xspace}
\newcommand{\delegator}{\textit{\textcolor{replygreen}{Delegator}}\xspace}
\newcommand{\delegators}{\textit{\textcolor{replygreen}{Delegators}}\xspace}
\newcommand{\learner}{\textit{\textcolor{inquiryred}{Learner}}\xspace}
\newcommand{\learners}{\textit{\textcolor{inquiryred}{Learners}}\xspace}
\begin{document}
\NewDocumentCommand{\cSay}{ o m }{%
  \textit{``#2''}% the quote
  % \IfValueT{#1}{%
  %   \hspace{0.4em}% adds breathing room
  %   {\footnotesize\textbf{#1}\,\raisebox{-0.2ex}{\scalebox{0.8}{\faComment}}}%
  % }%
}
%% Rights management information.  This information is sent to you
%% when you complete the rights form.  These commands have SAMPLE
%% values in them; it is your responsibility as an author to replace
%% the commands and values with those provided to you when you
%% complete the rights form.

%% Use this when submitting an article to a sponsored event. You'll
%% receive a unique submission ID from the organizers
%% of the event, and this ID should be used as the parameter to this command.
%%\acmSubmissionID{123-A56-BU3}

%%
%% For managing citations, it is recommended to use bibliography
%% files in BibTeX format.
%%
%% You can then either use BibTeX with the ACM-Reference-Format style,
%% or BibLaTeX with the acmnumeric or acmauthoryear sytles, that include
%% support for advanced citation of software artefact from the
%% biblatex-software package, also separately available on CTAN.
%%
%% Look at the sample-*-biblatex.tex files for templates showcasing
%% the biblatex styles.
%%

%%
%% The majority of ACM publications use numbered citations and
%% references.  The command \citestyle{authoryear} switches to the
%% If you are preparing content for an event
%% sponsored by ACM SIGGRAPH, you must use the "author year" style of
%% the next command will enable that style.

% Define custom comment commands
\newcommand{\yihung}[1]{\textcolor{blue}{[yihung: #1]}}   % for "V" reviewer
\newcommand{\jacob}[1]{\textcolor{teal}{[jacob: #1]}}   % for "V" reviewer
\newcommand{\tom}[1]{\textcolor{red}{[Tom: #1]}}    % for "O" reviewer
\newcommand{\jim}[1]{\textcolor{purple}{[Jim: #1]}} % general note
\newcommand{\vicki}[1]{\textcolor{orange}{[Vicki: #1]}} % general note
\newcommand{\boyuan}[1]{\textcolor{cyan}{[Gary: #1]}} % general note
\newcommand{\sarina}[1]{\textcolor{brown}{[Gary: #1]}} % general note
\newcommand{\krystal}[1]{\textcolor{olive}{[Gary: #1]}} % general note

% Allow author comments to be turned off
\newcommand{\includeAuthorComments}[1]
{
   \ifthenelse{\equal{#1}{0}}
   {
      \renewcommand{\yihung}[1]
      {
         {} % empty definition
      }
      \renewcommand{\jacob}[1]
      {
         {} % empty definition
      }
      \renewcommand{\tom}[1]
      {
         {} % empty definition
      }
      \renewcommand{\jim}[1]
      {
         {} % empty definition
      }
      \renewcommand{\vicki}[1]
      {
         {} % empty definition
      }
      \renewcommand{\boyuan}[1]
      {
         {} % empty definition
      }
      \renewcommand{\sarina}[1]
      {
         {} % empty definition
      }
      \renewcommand{\krystal}[1]
      {
         {} % empty definition
      }
   }{}
}

\author{Yi-Hung Chou}
\orcid{0000-0001-8537-0200}
\affiliation{%
  \institution{University of California, Irvine}
  \city{Irvine}
  \country{USA}
}
\email{yihungc1@uci.edu}

\author{June Hyung Kim}
\orcid{0009-0005-0820-9532}
\affiliation{%
  \institution{University of California, Irvine}
  \city{Irvine}
  \country{USA}
}
\email{jacobk13@uci.edu}

\author{Vasisht Kartik}
\orcid{0009-0003-8378-0486}
\affiliation{%
  \institution{University of California, Irvine}
  \city{Irvine}
  \country{USA}
}
\email{vkartik@uci.edu}

\author{Victoria Jackson}
\orcid{0000-0002-6326-931X}
\affiliation{%
  \institution{University of Southampton}
  \city{Southampton}
  \country{United Kingdom}
}
\email{v.jackson@soton.ac.uk}

\author{Tobias Schimmer}
\orcid{0009-0007-7728-0556}
\affiliation{%
  \institution{SAP Labs}
  \city{Newport Beach}
  \country{USA}
}
\email{tobias.schimmer@sap.com}

\author{Thomas Zimmermann}
\orcid{0000-0003-4905-1469}
\affiliation{%
  \institution{University of California, Irvine}
  \city{Irvine}
  \country{USA}
}
\email{tzimmer@uci.edu}

\author{James A. Jones}
\orcid{0000-0002-5359-7934}
\affiliation{%
  \institution{University of California, Irvine}
  \city{Irvine}
  \country{USA}
}
\email{jajones@uci.edu}
% Change the value to 0 (temporarily) if you want to preview what the length of the paper is with no author comments. Ultimately, we want to delete all author comments, but this is helpful to gauge how we are tracking in term of space.
\includeAuthorComments{1}
\title{Developers as Managers of AI Agents: A Workplace Shadowing Study}

\renewcommand{\shortauthors}{Chou et al.}

\begin{abstract}
As AI coding agents take on increasingly autonomous development work,
developers' roles are expanding from writing code to directing,
coordinating, and checking agents.
Yet little is known about how these activities unfold in everyday
professional software development.
We conducted a mixed-methods workplace shadowing study of 11
professional developers using AI coding agents on their own work.
Across approximately 15 hours of observation, we manually coded
developers' activities and task transitions, analyzed 255 submitted
prompts and their surrounding activities, and conducted  4.6 hours of follow-up interviews with all 11 participants.

We find that working with agents involved substantial management work.
Developers spent an average of 33.4\% of their observed time formulating
prompts or checking AI responses, while task-switching rates ranged from
0 to 101.6 switches per hour.
Their prompts served not only to produce software artifacts, but also to
understand agent behavior, clarify requirements, configure agents, and
coordinate work.
Connecting developers' prompting, checking, and task-management
practices revealed three recurring patterns:
\textit{Supervisor}, in which developers delegated work while directly
inspecting resulting artifacts;
\textit{Delegator}, in which developers coordinated agents through
dialogue and delegated both implementation and checking; and
\textit{Learner}, in which developers used agents primarily to build
understanding.
Developers relied on agents for technical information accessible from
code and documentation, while turning to colleagues for organizational
context, authority, and collective alignment.
Together, our findings characterize agentic development in the
workplaces we studied as a redistribution of developer control: from
producing artifacts toward directing delegated work, deciding what to
verify, and allocating attention across tasks.
\end{abstract}

%   <concept_id>10011007.10011074</concept_id>
%   <concept_desc>Software and its engineering~Software engineering</concept_desc>
%   <concept_id>10003120.10003121.10011748</concept_id>
%   <concept_desc>Human-centered computing~Empirical studies in HCI</concept_desc>
%   <concept_id>10003120.10003121.10003122.10011750</concept_id>
%   <concept_desc>Human-centered computing~Field studies</concept_desc>

% \ccsdesc[500]{Software and its engineering~Software engineering}
% \ccsdesc[300]{Human-centered computing~Empirical studies in HCI}
% \ccsdesc[300]{Human-centered computing~Field studies}

% \keywords{AI coding agents, developer work practices,
% workplace shadowing, human--AI interaction, agent management}

\maketitle

\section{Introduction}
Software engineering practice is being rapidly transformed by advances in large language models (LLMs).
AI-enabled development tools have progressed from code completion~\cite{ziegler2022productivity, barke_grounded_2023} to conversational coding assistants~\cite{lehtinen_lets_2024, 10.1145/3597503.3639187} and, more recently, coding agents capable of carrying out development tasks with increasing autonomy~\cite{10.1145/3797105, roychoudhury2026agentic}.
As developers delegate more work to these agents, their activities extend beyond writing code to directing agents, coordinating concurrent work, and deciding how to check generated outputs.
Yet these activities occur within the broader demands of professional software development, where developers must also manage ongoing tasks, information needs, tools, and collaboration.
Understanding how coding agents fit into this everyday work remains an open question.

Prior research has examined AI-assisted development through repository analyses~\cite{10.1145/3793302.3793349, murphyhill2026adoptionimpactcommandlineai}, surveys~\cite{liang_large-scale_2024}, observations of live-streamed programming~\cite{10.1145/3797105}, and controlled experiments~\cite{becker2025measuring}.
These studies showed a substantial code outputs from developers with AI agents~\cite{10.1145/3793302.3793349}, and higher perceived productivity and reduced workload~\cite{10.1145/3847122}. 
However, researchers have also reported substantial time spent waiting for generated outputs~\cite{10.1145/3797105}, less code quality~\cite{10.1145/3793302.3793349}, effort required to assess their correctness~\cite{gordon_co-audit_2023}, and, in some cases, longer task-completion times~\cite{becker2025measuring} and review time~\cite{11121707}. However, these perspectives provide a limited account of how professional developers incorporate coding agents into their everyday work.
Repository-level outcomes do not reveal the activities through which developers direct and check agents, while controlled tasks and publicly streamed sessions capture different conditions from workplace development~\cite{10.1145/3797105, alaboudi_exploratory_2019}.
Professional software development already requires developers to divide their attention across multiple tasks, artifacts, and collaborators~\cite{10.1145/1054972.1055017, mark2023attention, 10.1145/3210459.3210471}, while continually seeking answers to questions around artifacts, such as code and program behaviors~\cite{ko_information_2007, 4497212, 5710950}.
Developers must also coordinate dependencies and maintain awareness of other people and ongoing work~\cite{begel2009coordination, 5710950}, sometimes across globe and teams~\cite{al-ani_globally_2013, cataldo2008socio}.

Coding agents add another source and recipient of information in this environment.
Developers may now wait for agents while advancing other tasks, return later to inspect their progress or outputs, provide additional context, or seek information needed to decide what should happen next.
Understanding these interactions therefore requires examining how developers allocate attention and how information flows between developers, agents, artifacts, and collaborators during everyday work. To address this gap, we conducted a mixed-methods observational field study of 11 professional developers working with AI coding agents on their own development tasks in workplace settings during approximately 90-minute sessions.
We manually coded observable activities and task transitions, then linked prompts to preceding activities and subsequent checks.
Follow-up interviews used visualizations of participants' sessions to examine their interpretations of the observed patterns, checking strategies, and preferences for seeking information from agents versus colleagues, while allowing us to co-construct interpretations of the observed practices with participants~\cite{harvey2015beyond}.
We investigate two sets of research questions:

\begin{itemize}[nosep,leftmargin=*]
    \item \textbf{RQ1: Developers' Attention}
    \begin{itemize}[nosep,leftmargin=1em]
        \item \textbf{RQ1-1:} What activities do developers engage in, and how is their time distributed?
        \item \textbf{RQ1-2:} How frequently do developers switch tasks?
    \end{itemize}
    \item \textbf{RQ2: Information Flow Between Developers and Agents}
    \begin{itemize}[nosep,leftmargin=1em]
        \item \textbf{RQ2-1:} What information do developers observably seek from and provide to AI agents?
        \item \textbf{RQ2-2:} How do developers observably check agents' outputs?
        \item \textbf{RQ2-3:} What management patterns emerge from prompting and checking behaviors?
        \item \textbf{RQ2-4:} What information do developers prefer to obtain from agents versus human peers?
    \end{itemize}
\end{itemize}

RQs, pertaining to \textbf{Developers' Attention}, examines how developers allocate attention across activities and tasks while working with agents, building on prior work on AI-related waiting~\cite{becker2025measuring,10.1145/3797105} and context switching in software work~\cite{10.1145/1054972.1055017,10.1145/2635868.2635892,8666786,10.1145/3210459.3210471}.
The second set of RQs, pertaining to \textbf{Information Flow}, examines how developers interact with agents, and how these practices form recurring management patterns.
Together, the RQs shape how developers manage their AI agents in everyday   development~\cite{ko_information_2007}.

Our findings foreground the management work involved in using AI coding agents.
Prompt formulation and AI-response checking occupied an average of 33.4\% of observed session time.
Participants worked on 1--17 distinct tasks, with task-switching rates ranging from 0 to 101.6 switches per hour.
Developers not only requested development outputs but also clarified requirements, configured agent behavior, coordinated concurrent work, and decided what to check themselves or delegate back to agents.
Connecting preceding activities, prompt purposes, and subsequent checks revealed \textbf{three recurring patterns}:
(1)~\supervisor, combining work requests with direct artifact inspection;
(2)~\delegator, emphasizing clarification, agent configuration, and coordination through dialogue, often delegating both implementation and checking; and
(3)~\learner, seeking explanations to build understanding of unfamiliar code, tools, and data.
These patterns overlapped within sessions, describing ways of working rather than fixed developer types.
Participants further situated agent management within workplace collaboration, describing how they turned to colleagues for organizational context, human authority, and collective alignment, while using agents for technical information accessible from code, documentation, or public knowledge.

Our study makes two contributions: (1) an empirical account of how professional developers allocate attention, manage coding agents, and check their outputs in everyday work; and (2) three empirically derived
management patterns---\supervisor, \delegator, and \learner---that capture how developers combine artifact inspection, dialogue-based coordination, and explanation seeking.
\section{Background}
This section first reviews prior research on developers’ use of AI-assisted development tools (\S~\ref{subsec:DeveloperWithAI}), highlighting gaps in our understanding of developers’ productivity and work practices when using AI agents.
We then review broader observational research on developers’ everyday work practices (\S~\ref{subsec:ObserveDevAtWork}) before the release of AI, which informs the design of our study.

\subsection{Understanding Developers Working with AI\label{subsec:DeveloperWithAI}}
The rapid adoption and advancement of large language models (LLMs) have transformed how developers perform software engineering tasks. Prior research has examined the effects of AI-assisted development through repository mining, controlled experiments, and observations of publicly available programming sessions.

Through mining open-source GitHub repositories, Hao~\etal found that although AI-enabled development tools substantially increased the amount of code produced, their use could also reduce software quality as measured through static analysis~\cite{10.1145/3793302.3793349}. In a randomized controlled trial, Becker~\etal found that experienced open-source developers completed tasks 19\% more slowly when permitted to use AI tools; their analysis of screen recordings further showed that developers spent part of their implementation time prompting AI systems, waiting for generation, and reviewing generated outputs~\cite{becker2025measuring}. Similarly, through observations of livestreamed programming sessions, Chou~\etal found that all observed vibe coders spent more than 20\% of their session time waiting for AI-generated outputs~\cite{10.1145/3797105}. Huang~\etal observed 13 experienced developers and found that they maintained control over coding agents through planning, supervision, and output validation rather than fully delegating their work~\cite{huang2025professional}. Khojah~\etal collected ChatGPT histories of 24 professional developers over five consecutive workdays and qualitatively analyzed their dialogues, including the project-specific context developers provided and the purposes for which they sought assistance~\cite{10.1145/3660788}.

However, these studies examined AI-assisted development in settings that differ from ongoing workplace activities. Becker~\etal focus their study around predefined repository issues and measured developers' completion time at the individual-issue level~\cite{becker2025measuring}, whereas Huang~\etal observed each participant performing a single self-selected task during a 45-minute think-aloud session~\cite{huang2025professional}. These designs provided limited insight into how developers transition among multiple concurrent workplace activities and coordinate their work across agents, tools and colleagues. Chou~\etal examined livestreamers, whose behaviors may have been shaped by the need to entertain their audience~\cite{10.1145/3797105, alaboudi_exploratory_2019}. Despite the growing body of research on AI-assisted development, close examinations of how developers work with AI coding agents in their actual workplaces remain limited. To address this gap, we conducted an observational study of professional developers in their natural work settings. Our methodological design draws on prior observational studies in software engineering, which we review next.

\subsection{Observing Developers at Work\label{subsec:ObserveDevAtWork}}

Because direct observations of developers using AI coding agents in professional workplaces remain rare, we primarily draw methodological inspiration from observational studies conducted before the emergence of agentic software development.
These studies have examined developers' work in situ through think-aloud protocols, direct observation, field notes, interaction traces, and surveys, providing established approaches for understanding software work in its organizational context.

Ko~\etal conducted a 90-minute observational study with a think-aloud protocol to characterize developers' information needs in collocated software development~\cite{ko_information_2007}.
They identified recurring unmet information needs, many concerning development artifacts such as source code and program state.
This work motivates our examination of the information exchanged between developers and coding agents. A complementary line of research has examined how developers allocate attention across everyday work.
Meyer~\etal, through observational and survey studies, documented the fragmented nature of developers' work and frequent transitions among activities~\cite{10.1145/2635868.2635892, 7829407, 8666786}.
Chong and Siino similarly studied developers in their workplace, but relied on extended direct observation and field notes: over seven months, they accumulated more than 40 hours of observations and analyzed 242 interruptions~\cite{10.1145/1180875.1180882}.
More broadly, Mark~\etal's research on workplace attention has shown increasingly short periods of engagement with individual screens or applications~\cite{10.1145/1357054.1357072, mark2023attention, 10.1145/1054972.1055017}.
These findings motivate our examination of how developers divide attention across tasks while working with coding agents, particularly when agent work can continue in parallel.

Drawing on these traditions, our study combines direct observation, field notes, video recordings, and follow-up interviews.
We avoid a concurrent think-aloud protocol to reduce interference with participants' ongoing development work and the additional burden of participation.
Because interpreting screen recordings can nevertheless require some inference about task boundaries and observed practices, we use follow-up interviews to present participants with visualizations of their sessions, check whether the observed patterns reflect their typical work, and refine our interpretations of those patterns.
\section{Method\label{sec:method}}
\begin{figure*}[t]
\centering
\resizebox{\linewidth}{!}{%
\begin{tikzpicture}[
  stagebox/.style={
    draw, rounded corners=2pt, align=center,
    text width=2.35cm, minimum height=0.85cm,
    inner sep=3pt, font=\footnotesize
  },
  arrow/.style={-{Latex[length=2mm]}, thick}
]
\node[stagebox] (r)  at (0,0)    {Recruitment\\Two sites};
\node[stagebox] (o)  at (2.9,0)  {Phase 1\\Observation};
\node[stagebox] (a1) at (5.8,0)
  {Phase 1 Analysis\\[-1pt]
   {Preliminary Results}};
\node[stagebox] (i)  at (8.7,0)  {Phase 2\\Follow-up interviews};
\node[stagebox] (a2) at (11.6,0) {Phase 2\\Analysis};

\draw[arrow] (r) -- (o);
\draw[arrow] (o) -- (a1);
\draw[arrow] (a1) -- (i);
\draw[arrow] (i) -- (a2);
\end{tikzpicture}%
}
\caption{Study procedure. Interviewees reflected on  visualizations derived from observations.}
\vspace{-10px}
\label{fig:study-design}
\end{figure*}
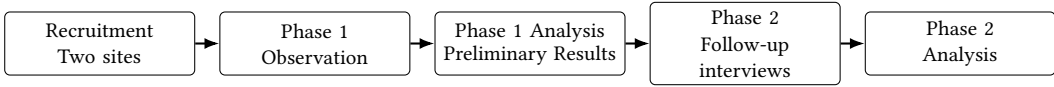
We conducted a two-phase mixed-methods observational field study of professional developers using AI agents on their own work at their workplace to strengthen ecological validity. The first phase combined quantitative summaries of manually coded activity traces with qualitative analysis of prompts and their surrounding activities. After completing the observational analysis, we conducted follow-up interviews where participants reflected on visualizations of their own sessions and anonymized patterns from other participants.

\subsection{Recruitment and Participants}
We recruited participants from two companies: Company A is a multi-national enterprise software company and Company B is a foundation that organizes the development of an open-source project that has more than 1,000 developers.
Authors 1 and 2 worked as contractors at Company A; Author 1 had previously interned there. 
Their roles enabled in-person recruitment at the company's U.S. office. 
To include developers working in a different organizational context, Author 2 recruited two developers through a personal connection at Company B (P7, P8). 
The authors remained subject to applicable confidentiality agreements, and all participants consented under the protocol approved by our Institutional Review Board.

Participants were eligible if they were at least 18 years old, located in the United States, worked professionally in software development, and used an AI coding agent. 
We recorded sessions with 13 developers and included 11 in the analysis. 
We excluded one session as an unscheduled meeting arose during the observation period, leaving insufficient time to observe development work.
We excluded another as the participant reached their AI coding agent's token limit and could not use the agent during the session. 
Table~\ref{tab:participants}
summarizes the backgrounds of the 11 included participants.

\begin{table}[htbp]
\centering
\caption{Participant overview. \textit{Obs.\ dur.}\ = duration of the recorded coding session.
\textit{Int.\ dur.}\ = duration of the post-session reflection interview.
Task-type codes: \textsc{fd}~=~Feature~Development, \textsc{cr}~=~Code~Review, \textsc{lnr}~=~Learning~, \textsc{m}~=~Maintenance}
\label{tab:participants}
\footnotesize
\begin{tabular}{llrllrrll}
\toprule
\textbf{P\#} &
\textbf{Tool(s)} &
\textbf{Prof.\ tenure} &
\textbf{Tasks} &
\textbf{Task types} &
\textbf{\# Prompts} &
\textbf{\# Events} &
\textbf{Obs.\ dur.} &
\textbf{Int.\ dur.} \\
\midrule
P1  & Claude Code & 7--10 yrs &  4 & FD, CR         & 12 & 343 & 1\,hr 17\,min & 17 mins \\
P2  & Claude Code & 4--6 yrs  &  4 & FD, CR         & 16 & 670 & 1\,hr 26\,min & 24 mins \\
P3  & Claude Code & <1 yr     &  1 & LNR            & 34 & 374 & 1\,hr 30\,min & 30 mins \\
P4  & Claude Code & 4--6 yrs  &  7 & FD, LNR, M     & 16 & 450 & 1\,hr 28\,min & 17 mins \\
P5  & Claude Code & 4--6 yrs  &  8 & CR, M          &  5 & 464 & 1\,hr 26\,min & 22 mins \\
P6  & Claude Code & 7--10 yrs &  5 & FD, M          & 49 & 381 & 1\,hr 15\,min & 31 mins \\
P7  & Codex       & 7--10 yrs & 13 & FD, CR         & 39 & 417 & 1\,hr 23\,min & 41 mins \\
P8  & Codex       & 4--6 yrs  & 17 & FD, CR         & 42 & 344 & 1\,hr 20\,min & 44 mins \\
P9  & Claude Code & 7--10 yrs &  2 & FD, LNR        & 25 & 348 & 1\,hr 11\,min & 22 mins \\
P10 & Claude Code & 4--6 yrs      &  2 & FD, LNR        & 17 & 327 & 1\,hr 24\,min & 12 mins \\
P11 & Claude Code & 7--10 yrs &  2 & FD, LNR        & 17 & 363 & 1\,hr 11\,min & 17 mins \\
\bottomrule
\end{tabular}
\end{table}

\vspace{-5px}
\subsection{Phase 1: Observational Study \& Preliminary Analysis}

\subsubsection{Observational Study} We asked each participant to reserve a 90-minute period free of meetings and to work on their own development activities rather than a researcher-assigned task. We intentionally left ``development work'' open to participants' interpretation. The resulting sessions included feature development, maintenance, code review, and exploration of unfamiliar codebases, as summarized in Table~\ref{tab:participants}.

\textbf{Briefing (5--10 min): }Each session began with a briefing during which the researcher
briefed the study procedure, captured consent and recording setup, and answered participants' questions. 

\textbf{Observation (70--90 min): }
A researcher remained present throughout each session, either
in person or remotely (P7, P8). All sessions were recorded
through videoconferencing software: Microsoft Teams for
in-person sessions and Google Meet for remote ones.
We did not employ a think-aloud protocol because continually
verbalizing thoughts could disrupt participants' workflows.

During the sessions, the researcher recorded field notes about
activities that could not be interpreted from the screen alone,
such as stepping away, using a phone, or speaking with a
colleague. Because recording software recorded only one screen
at a time, we asked in-person participants to conduct most
of their work on a single screen. They could use another
screen for private activities, such as messaging; the researcher
noted the activity and its timestamp without recording its
content. Remote participants were asked to work on a single
screen and notify the researcher when they stepped away or
were otherwise inactive. These field notes contextualized
the screen recordings during analysis by documenting
off-screen activities.

\textbf{Optional Interview (5--10 min): } The onsite researcher reserved an optional short interview  at the end of each session, conducted either in-person or remotely, to clarify task contexts or observed activities that could not be readily interpreted from the recording. 
Across the 11 participants, the resulting observational dataset contains around 15 hours, 4,481 activity segments (see below), 280 prompt-formulation episodes, of which 272 were submitted.
\subsubsection{Phase 1 Data Analysis\label{subsubsec:analysis}}
\paragraph{RQ1: Activities, time allocation, and context switching.}

To understand how developers devoted their time, we manually
divided each recording into activity segments. 
We adapted the activity coding scheme from
Chou~\etal~\cite{10.1145/3797105}, a recent study of vibe coding practices that similarly analyzed video-recorded development sessions. Because screen recordings do not reveal participants'
attention or cognitive states, we coded only observable actions
and artifacts, using labels such as \emph{Check AI Response}
rather than \emph{Read AI Response}.
We adapted the initial codebook from the same study, added or refined codes as new activities emerged, and revisited previously coded recordings whenever definitions changed.
Each recording was coded in full by one researcher.
A second researcher then watched the entire recording and reviewed all segment boundaries and assigned codes, identifying potential corrections and ambiguous cases.

During weekly meetings, researchers discussed proposed corrections and ambiguous cases identified during review, including disagreements about segment boundaries (i.e., \texttt{Wait} or \texttt{Check AI}) and the observable evidence required to assign a code. They reached consensus on the final coding and refined the codebook when needed. 
For example, researchers clarified the minimum duration of observable inactivity required to operationally label a segment as \emph{Wait}.

\begin{wraptable}{r}{0.55\textwidth}
\vspace{-5px}
\centering
\small
\caption{A coding snippet from P1 spanning the first three prompts. P1 has multiple Claude Code sessions opened across code editors and terminals. Pr1C refers to the activities used to verify the resulting artifacts,  outputs, or AI response, PrXC refers to the checks of the prompts that occurred before the video session }
\label{tab:coding-snippet}

\begin{tabularx}{\linewidth}{@{}llllX@{}}
\toprule
Time & Task & Window & PID & Codebook category \\
\midrule
3:17 & A  & \textsf{terminalA} &              & \textsc{Window Switch} \\
3:18 & A  & \textsf{terminalA} & \texttt{Pr1}  & \textsc{Prompt Formulation} \\
3:47 & B  & \textsf{editorB}   & \texttt{PrXC}  & \textsc{Check Code} \\
3:59 & B  & \textsf{editorB}   & \texttt{PrXC} & \textsc{Check Code Diff} \\
4:16 & -- & --                &              & \textsc{Window Switch} \\
4:17 & B  & \textsf{terminalB} & \texttt{PrXC} & \textsc{Check AI} \\
4:23 & -- & --                &              & \textsc{Window Switch} \\
4:34 & B  & \textsf{editorB}   & \texttt{Pr2}  & \textsc{Prompt Formulation} \\
5:43 & B  & \textsf{editorB}   & \texttt{Pr2C} & \textsc{Check AI} \\
5:44 & B  & \textsf{editorB}   & \texttt{Pr2C} & \textsc{Approve AI Action} \\
5:45 & -- & --                &              & \textsc{Window Switch} \\
5:47 & C  & \textsf{terminalC} & \texttt{PrXC} & \textsc{Check AI} \\
5:53 & C  & \textsf{terminalC} & \texttt{Pr3}  & \textsc{Prompt Formulation} \\
\bottomrule
\vspace{-15px}
\end{tabularx}
\end{wraptable}

Such an abductive qualitative coding approach yielded new dimensions that were not previously captured and allowed us to refine our coding scheme iteratively. 
For example, unlike Chou~\etal~\cite{10.1145/3797105}, where the live-streamers often focus on a single task, we found our participants frequently switching between different independent tasks. Therefore, we associated each activity segment with a task through the ``git branch'' identifier visible in all participants' code editors, or AI-agent sessions.
When these identifiers were unavailable (for example, while viewing documentation, a GitHub page, or a database inspection tool), we used surrounding temporal context to infer the associated task. 
When ambiguous, we asked the participant immediately after the session as discussed in the previous section.

Table~\ref{tab:coding-snippet} illustrates an example of resulting annotations, including how activities were associated with tasks, windows, and prompts. We counted a task transition whenever consecutive task-related activities were assigned to different tasks, including brief checks of another task. For example, Task~A $\rightarrow$ Task~B $\rightarrow$ Task~A represents two transitions involving two unique tasks. From these annotations, we calculated each participant's proportion of recorded time spent on each activity and the frequency of task transitions.

\paragraph{RQ2: Information between Agents and Developers}
We used each sent prompt as an analytical anchor to examine the observable practices of developers, both (RQ2-1) \emph{before their prompt} to examine the information that informs their prompts, and (RQ2-2) \emph{after their prompt} to examine the information provided by the agent's output.

For RQ2-1, we conducted a \textit{backward analysis}. We analyzed each prompt's content and reviewed the coded activities during the ten minutes before the developer began formulating it. This interval captured preceding activities such as inspecting source code, documentation, tickets, or previous agent responses. When visible evidence suggested that information included in the prompt had been obtained earlier (\eg a paste from an earlier copy action), coders extended their review beyond the initial ten-minute window.

For each prompt, we coded three aspects: (1)~its \emph{intent} (inquiry, request, or reply); (2)~the \emph{type of context explicitly copied and pasted into it}, such as documentation, source code, or terminal output; and (3)~the \emph{recorded on-screen activities that provided evidence of where the pasted context originated}, (4)~\emph{the recorded preceding activities} without assuming that every inspected source contributed information to the prompt.
For example, P4 spent the minutes before one prompt reading a ticket, the associated GitHub PR page, and an internal wiki; they then pasted the wiki content directly into an inquiry asking the agent to explain what changes were still required for a migration with external dependency.
Here we could trace a clear chain from the embedded context (\emph{pasted documentation}) back to its observed source (the wiki). Prompts could receive multiple intent labels. Our coding thus rested on two kinds of observable evidence: context the developer explicitly pasted into the prompt, and activities visible on screen shortly before it. Information that a developer may have drawn on from their memory, or from sources consulted outside the recorded session, were not observable and therefore fell outside our analysis.

For RQ2-2, we conducted \textit{forward analysis} by treating each submitted prompt as the unit of analysis and traced
the recording forward to identify all subsequent activities, such as
\emph{Check AI Response} or \emph{Interaction with Terminal}, that were potentially associated with the prompt, stopping at the next prompt within the
same task. 
For each prompt, we recorded the number of \emph{distinct types} of
``checking'' it received, and coded the checking practices performed on its resulting output. 
These practices spanned both inspecting the \emph{content} of the output (\eg checking generated code or the agent's
response) and exercising it (\eg running the program, tests, or a command and examining the resulting output). 
A prompt could therefore receive multiple
checking labels when a participant used more than one type of practice.
A follow-up prompt could itself
serve to verify the prior turn. For instance, asking the agent ``are you
sure?'', which we coded as a \emph{follow-up}. 
We traced checking activities until the next submitted prompt within the same task, or the end of the recording if no subsequent prompt occurred in that task. Checks occurring after that subsequent prompt were not linked back to the earlier output. If no checking activity
was observed within this interval, we labeled the prompt
\emph{No Check Observed}. Response checks and verification activities were not necessarily adjacent to the corresponding prompt. Participants could switch to another task while an
agent was working and later return to check or verify its output, as illustrated
in Figure~\ref{fig:interleaved}. We therefore traced prompt-related activities
across intervening tasks and windows rather than restricting the analysis to
activities immediately following prompt submission.

% \vicki{so what did you capture as a result of the analysis? This is unclear. Was it the artifacts used to inform the prompt's content? Did you capture unclear when it was unclear, perhaps because of tacit knowledge from the developer?}\yihung{I will add the example episode to explain.}

\subsection{Phase 2: Follow-up Interviews}
Only after completing the observational coding and preliminary analysis did we invite all participants to a 20-minute follow-up interview. The final dataset contains interviews with the same 11 participants from the observational study. While initial interviews were scheduled for 20 minutes, sessions were allowed to extend up to 44 minutes when participants offered rich and unsolicited accounts and insights that required further exploration. This flexible approach ensured thorough data collection while prioritizing participant time.
Informed by Harvey's~\cite{harvey2015beyond} dialogic approach, we used the follow-up interviews for member-checking and to develop interpretations of the observed patterns through discussion with them.
The interviews followed a common semi-structured guide supplemented by participant-specific questions derived from each recording.
Each topic was accompanied by a visualization of the relevant observations. 
The interviewers asked whether the activity segment  (Fig.~\ref{fig:time-allocation}) and task-switching pattern (Fig.~\ref{fig:interleaved}) were representative of participants' usual work, elicited explanations for patterns and surprising moments to them, and provided perspectives not directly observable in the recordings.
For example, when discussing task switching, participants viewed their own task-transition pattern alongside anonymized patterns from other participants.

We analyzed the interview responses primarily through deductive
coding, using the guide's topics: time allocation, task
switching, documentation, collaboration, and verification practices, as predefined categories. Within these categories,
we developed inductive subcodes to capture distinctions in
participants' accounts.
For RQ1, these accounts helped explain time allocations and task switching in relation to tasks, responsibilities, and perceived effects of agents on focus. For RQ2, they helped interpret observed information-seeking and agent-management practices by providing participants’ perspectives on information-source preferences and checking strategies.

\subsection{Threats to Validity}

\paragraph{Credibility.}
Participants may have altered their behavior because a researcher was present and their screens were recorded. We asked them to work on their own development activities rather than assigned tasks, but cannot determine whether they would have prompted agents, checked outputs, or switched tasks in the same way without observation. We used \textbf{member checking} in the follow-up interviews. Their responses helped contextualize the recordings but cannot rule out an observer effect~\cite{kitchenham2002preliminary}. The recordings capture visible actions rather than participants' attention or intentions. For example, P6, P7, and P8 frequently switched between agent sessions, but we cannot distinguish whether they read an agent's output or briefly checked whether it needed input. We therefore coded observable actions without inferring the depth of participants' engagement. One researcher coded each recording, a second watched and reviewed it in full, and the team discussed ambiguous cases. We also maintained \textbf{reflexive notes} documenting our perspectives and used \textbf{peer debriefing}~\cite{creswell_qualitative_2016} with three software-engineering professors to examine our interpretations.

\paragraph{Dependability.}
Our time allocations and task-transition counts depend on how observable activities were segmented and assigned to tasks. Under our stated definition, even a brief visit to another task counts as a transition. The counts thus describe visible movement between tasks, not necessarily substantial changes in work focus. Readers should interpret comparisons in light of this definition, particularly for participants who made frequent brief visits to other tasks.

\paragraph{Transferability.}
We observed 11 professional developers working on their own tasks for approximately 90 minutes each. The participants were recruited through the authors' access to two workplaces; their projects, organizations, and tools may differ from those of other developers. Sessions were scheduled to be free of meetings and thus do not represent every part of a workday. The 11 sessions cannot capture the full range of ways developers work with AI coding agents. As P6 noted: \cSay{[when looking at developers with few tasks], you maybe have interviewed the wrong people or at the wrong time, ...[busy developers]...you may never actually get to interview those people because those people be like, I don't have time for you}. We therefore present the findings as practices observed in these work contexts, not as an exhaustive account or estimates of their prevalence among professional developers.

\paragraph{Limits of recordings and retrospective accounts.}
Video recording software recorded one screen at a time. P1, P4, and P9 were the only participants using two monitors during recording. They could move private messaging to the unrecorded monitor if they did not want its content captured. When they did so, the researcher noted that messaging occurred and recorded its timing in field notes, without documenting the message content. We therefore could not determine whether an off-screen message informed a later prompt. All 11 participants took part in follow-up interviews between one week and three months after observation. This interval may have limited recall of specific moments. We anchored questions in visualizations of their recorded activities and asked primarily about their general workflows. We treat their explanations as retrospective perspectives rather than evidence of what they were thinking during each recorded action.

\paragraph{Ethical considerations.}
The study followed an approved Institutional Review Board protocol, and participants consented to observation and recording. 
One member was employed at Company A but did not supervise participants, conduct recruitment, or have access to participant identities or recordings. 
Participants were informed that participation was voluntary and that results shared beyond the first three authors, who coded the recordings, would be anonymized. 
The researchers' workplace access was governed by the applicable study and confidentiality arrangements. Although some messages with colleagues were visible in the recordings, we did not analyze their content to protect participants' and colleagues' privacy. We therefore report time spent on communication but do not characterize the information exchanged in those messages.

\section{Results}
We answer the research questions by bringing together observational videos
(\faIcon{eye}) and follow-up interviews
(\faIcon{comment}). The observational videos ground our account in what developers did during the sessions; the interviews capture how developers interpreted those activities after seeing the observational findings and how they related them to their usual practices. 
We draw on both sources for RQ1-1 through RQ2-3. RQ2-4 draws primarily on participants’ retrospective reflections.
Throughout the results, terms such as ``checking'' describe actions coded from screen recordings and do not establish participants' gaze, attention, or thought processes, which we did not capture through eye tracking or think-aloud data.
\begin{figure}[h]
    \vspace{-10px}
    \centering    \includegraphics[width=1\linewidth]{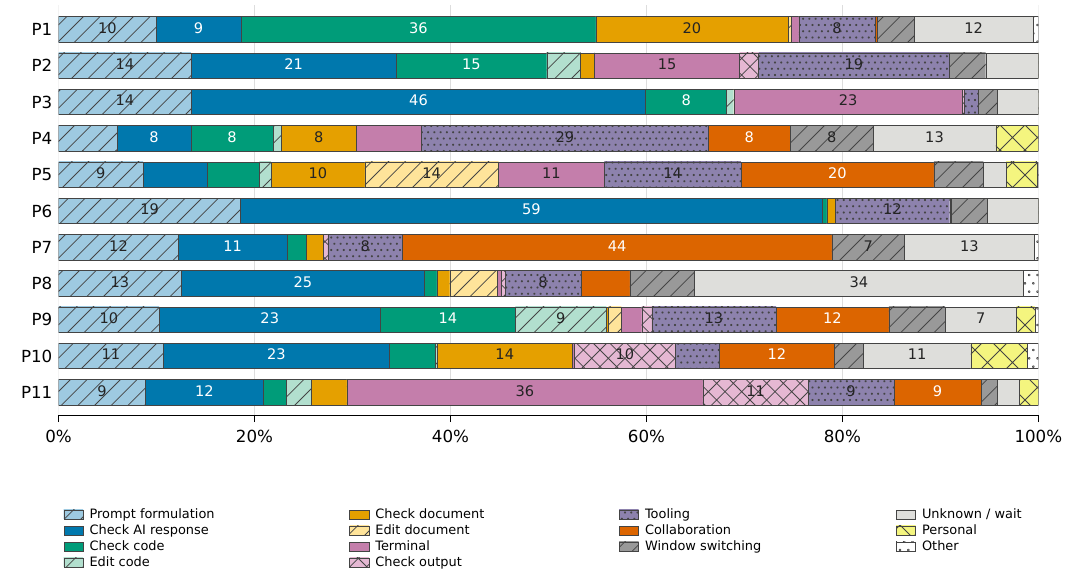}
    \caption{Distribution of recorded session time across activities for each participant.}
    \label{fig:time-allocation}
    \vspace{-14px}
\end{figure}
\subsection{RQ1-1: Time Allocations Among Activities}
\textbf{Finding 1 (\faIcon{eye}): Agent interaction occupied a substantial
share of session time}, with considerable variation across
developers. We segmented the videos by activity and aggregated the time
spent on each activity (Fig.~\ref{fig:time-allocation}).
Participants spent an average of 33.4\% of their sessions
on AI-related activities: 11.3\% formulating prompts
(range: 6.0\%--18.6\%) and 22.1\% checking AI responses
(range: 6.6\%--59.4\%). They spent 8.9\% of their sessions inspecting code
(range: 0.5\%--36.2\%), 5.4\% checking documents, and
8.7\% interacting with the terminal. Individual variation
was substantial, consistent with
Chou~\etal~\cite{10.1145/3797105}.
P1 spent 36.2\% of their session inspecting code, whereas
P6, P7, P8, and P11 each spent less than 3\% doing so
(0.5\%, 2.0\%, 1.3\%, and 2.4\%, respectively).
P6 spent 77.9\% of their session formulating prompts or
checking AI responses. For ten of the eleven participants,
these agent-related activities occupied more time than
direct code inspection.

\textbf{Finding 2 (\faIcon{eye}): Managing windows accounted for a measurable share of developers' session time.}
% \faIcon{eye}
Window switching---including moving or arranging windows
and switching between applications or macOS workspaces---accounted
for an average of \textbf{4.6\%} of session time
(range: 1.6\%--8.5\%). This represented approximately one
to eight minutes per session and roughly half the average
time spent viewing or editing code files (10.6\%).

All participants confirmed the distributions broadly matched their workdays at the time of recording. Some participants (P3, P5, P8, P9) had not previously estimated their own time allocation: P3 admitted \cSay{I guess I actually never thought about it that much}. Some were surprised by the time spent on prompting (P2, P3), documentation (P5), or window switching (P4, P9). P2 reflected: \cSay{the prompting of AI took so much of the time...but...I guess it's plausible...you want to be really specific in the problem.} P4, when showing the window switches pattern (8.5\%), they stated: \cSay{I work on multiple stuff at a time, so I try to be very organized}

\textbf{Finding 3 (\faIcon{comment}): Participants explained their time allocation through role (P1--P8), organizational context (P8), and team norms (P6, P7, P8).} 
P3 described how their work had shifted away from writing code: \cSay{I not really write much code anymore...my work very switched a lot from actively writing code to reviewing code and thinking about the architecture.} Many participants (P2, P3, P7, P8) also highlight the reduced time for coding.
P4 attributed their broad task distribution to their dedicated DevOps responsibilities, while P5 tied their high documentation time to their role expectations. P7, working on an open-source project, explained their high messaging and collaboration time through team structure. P6, P7, and P8 further described agent-first team norms to account for their limited code-inspection time: P6 noted \cSay{we don't actually review...you just click override, merge, done}; P7 similarly described a culture of trusting agent output---\cSay{if the agent thinks this is safe to land, then let it land}---while P8 reported \cSay{we self-merge} and noted having unconstrained tokens. Some participants also noted that their practices had since changed: P6 reported spending \cSay{a lot less time reviewing whatever the AI is doing}, whereas P2 said \cSay{reviewing the AI code maybe got even more.}

\begin{figure}[t]
\centering
\includegraphics[width=0.9\linewidth]{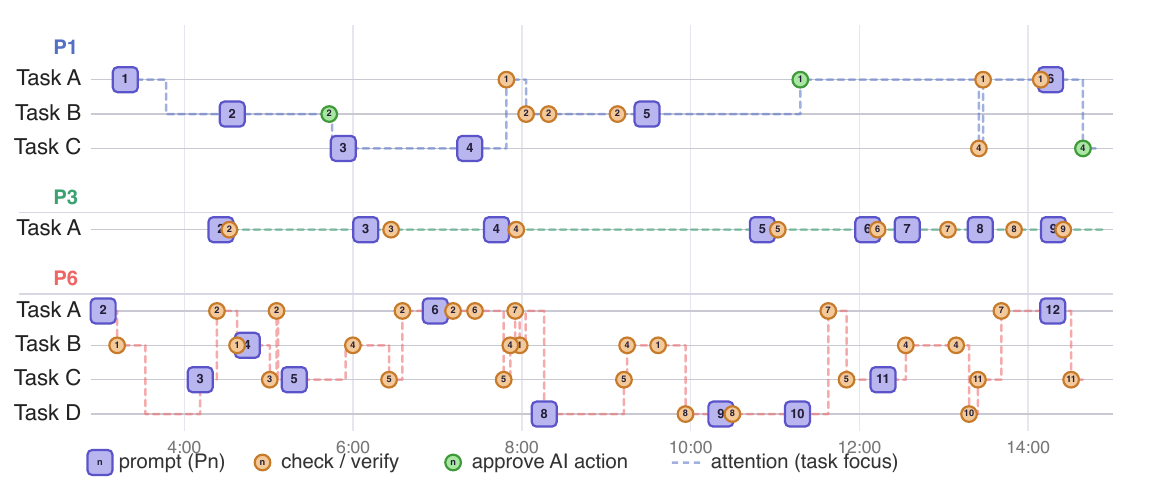}
\caption{Interleaved prompting and deferred output checking during a 12-minute
excerpt from P1, P3 and P6's session. Each swimlane represents an anonymized task.
Squares denote submitted prompts, and circles denote observable monitoring,
checking, or approval activities associated with their outputs. Matching
indices connect prompts to their associated activities when the relationship
was supported by the recording. The dashed line traces the participant's coded
active task over time. \label{fig:interleaved}
}
\vspace{-6pt} 
\end{figure}

\subsection{RQ1-2: Switches Among Tasks}
Given that developers spent an average of 4.6\% of session time switching or arranging windows, we next examined task switching
using the task labels described in \S~3.2.2.

\begin{wraptable}{r}{0.45\textwidth}
\centering
\footnotesize
\setlength{\tabcolsep}{3pt} % default: 6pt
\caption{Task-switching statistics per participant. Sw.\ = switches;
Sw/hr = switch rate per hour; $\bar{x}$, $\tilde{x}$ = mean and median run duration.}
\label{tab:task-switching}
\begin{tabular}{@{}lrrrrrrr@{}}
\toprule
\textbf{P\#} &
\textbf{Tasks} &
\textbf{Sw.} &
\textbf{Sw/hr} &
\multicolumn{4}{c}{\textbf{Run duration (min)}} \\
\cmidrule(l){5-8}
& & & & \textbf{min} & \textbf{$\bar{x}$} & \textbf{$\tilde{x}$} & \textbf{max} \\
\midrule
P1  &  4 &  33 &  25.76 &  0.03 &  2.10 &  0.96 & 27.10 \\
P2  &  4 &  31 &  21.66 &  0.03 &  2.68 &  0.55 & 20.97 \\
P3  &  1 &   0 &   0.00 & 89.62 & 89.62 & 89.62 & 89.62 \\
P4  &  7 &  30 &  21.14 &  0.02 &  2.56 &  1.45 & 16.47 \\
P5  &  8 &  17 &  11.91 &  0.17 &  4.61 &  0.93 & 33.28 \\
P6  &  5 & 124 & 101.60 &  0.02 &  0.56 &  0.32 &  3.13 \\
P7  & 13 &  36 &  26.59 &  0.08 &  1.62 &  0.72 & 14.62 \\
P8  & 17 &  79 &  62.73 &  0.02 &  0.73 &  0.21 &  8.40 \\
P9  &  2 &  25 &  21.21 &  0.01 &  2.61 &  0.51 & 17.89 \\
P10 &  2 &   1 &   0.75 &  1.84 & 40.25 & 40.25 & 78.66 \\
P11 &  2 &  21 &  17.96 &  0.04 &  2.97 &  1.82 & 15.34 \\
\bottomrule
\end{tabular}
\end{wraptable}

\textbf{Finding 4 (\faIcon{eye}): Task-switching patterns ranged from
single-task focus to over 100 switches per hour.}
Participants worked on 1--17 distinct tasks, with switching
ranging from none (P3) to 101.6 switches per hour (P6).
Task count did not necessarily correspond to switching frequency:
P6 switched among five tasks more frequently than P8 did among
17. Nine participants had median task runs shorter than three
minutes, but brief runs could coexist with sustained work.
For example, P5 had a median run of 0.93 minutes and a longest
run of 33.28 minutes.
Table~\ref{tab:task-switching} summarizes
task switching statistics, while Figure~\ref{fig:interleaved}
illustrates how task changes interleaved with prompting and
reviewing agent outputs. We selected these three sessions to illustrate contrasting
task-switching patterns: P1 alternated among a small set of
tasks, P3 remained on a single task, and P6 switched frequently
among five tasks. The contrast between P6 and P7--P8, who
worked on 13 and 17 distinct tasks, respectively, further
shows that task count and switching frequency capture
different aspects of task interleaving. Visualizations for
all 11 participants are available in the replication package. 

These patterns show how often developers moved between tasks, but do not establish what drove their switching or how AI influenced it. We therefore presented developers with visualizations of their own sessions and asked whether the patterns were typical and what drove them. If developers did not spontaneously mention AI, we explicitly asked how it affected their task switching. All developers reported that the visualizations reflected their typical level of task switching (all P). P1 remarked, \cSay{I'm so surprised to see [my typical work] pattern recognized using this visualization, typically at most, I have 3 different work going on.}

\textbf{Finding 5 (\faIcon{comment}): Developers attributed task switching to mental load, work responsibilities, agent orchestration, and waiting for CI/CD blockers (P8).} Participants described mental load as shaping how they structured their work (P3, P7, P8). P3 deliberately limited scope to a single task to avoid managing multiple pipeline stages: \cSay{I kind of focus on one thing and do not have to think about multiple stages of the pipeline.} P7 observed that while agents expanded active work items, parallel attention has limits: \cSay{you just can't work on that many things — even [thirteen] is more than the human brain could actually do effectively.} P8 described a growing fatigue from managing too many concurrent tasks: \cSay{it's just...AI fatigue...[with] so many tasks going...it's a different learning curve.} Work responsibilities, particularly cross-team coordination, also drove switching (P2, P10). P10 was simultaneously involved in two sub-teams with separate workstreams, creating unplanned context switches. P2 described inheriting support requests from former projects after moving to new role: \cSay{[it] ends up on our table because they're like oh you're the expert, you should take a look.}

Several participants attributed task switching to agent orchestration: dispatching long-running tasks and switching to other work while waiting for results (P2, P7, P8, P9). P2 described parallelizing work alongside Claude, though noted it could become \cSay{a context switching hell.} P7 noted that agents allowed many items to run in the background with only a few requiring active attention: \cSay{I can work on like fourteen distinct things during a day but it's really like only three of them are like in my brain, the other ones just happen in the background.} P8 observed that \cSay{the agent is doing the work and so I'm sitting back...my mind wanders...and so I end up doing a lot of things more shallowly,} while P9 described setting up separate AI sessions per task to manage the concurrent workload. 

% Finally, participants attributed switching to waiting for external blockers (P8). P8 identified CI verification and colleague decisions as the primary forcing functions: \cSay{CI...verification is just [eighty] percent of my blocking, so I [have to] bounce around,} and noted that delays also came from being \cSay{blocked by someone else's decision.}

\textbf{Finding 6 (\faIcon{comment}): Developers reported that agents supported focus \textit{within} tasks but complicated attention management \textit{across} tasks.}
When discussing agents' effects on focus, participants described benefits for within-task concentration (P2, P3, P4, P7, P9), including reduced effort in seeking information across sources (P2, P9), easier re-entry into tasks through conversation history (P4), and deeper engagement on hard or unfamiliar work (P3, P7), alongside reported difficulties managing attention across parallel agent tasks (P1--P3, P5--P8).
P2 contrasted earlier information-seeking with agent use: \cSay{You had to Google, research on different websites... Now you just use deep research on Claude... [it] makes it much more easy to stay focused.} P4 noted that conversation history eased re-entry after switching: \cSay{When I come back to it, I see the chain of responses... [it] helps me regain focus.}
In contrast, P8 described agents filling waiting time with other activities, leading to doing \cSay{a lot of things... more shallowly than focusing on one,} while P1 reported a persistent mental overhead of tracking parallel tasks: \cSay{in the back of my mind, I'm like, oh, did that review process finish, then I have to switch back.}
Participants also described strategies for managing agents. P11 and P1 reported deliberate task switching: \cSay{I purposely designed what to switch. (P11)} P6 used watch timers to revisit parallel tasks: \cSay{[I] set [myself] 10-minute timers because I forget to check in on certain results of my tasks.} P9 maintained a personal notes repository to preserve session context across agent tasks.

\subsection{RQ2-1: Directing and Informing AI Agents}
\label{sub:InfoToAI}

Through \textit{backward analysis} described in \S\ref{subsubsec:analysis}, we examined 280 prompt-formulation episodes. We excluded eight unsent or interrupted episodes and 17 agent commands unrelated to development or information seeking (\eg \texttt{\textbackslash exit}), retaining 255 submitted prompts for analysis.
\begin{table}[t]
    \centering
    \caption{Observable activities preceding prompts, by prompt type.
Cells show mean within-participant percentages
(prompt count $\cdot$ participant count).\label{tab:before-to-prompt}}
    \includegraphics[width=1\linewidth]{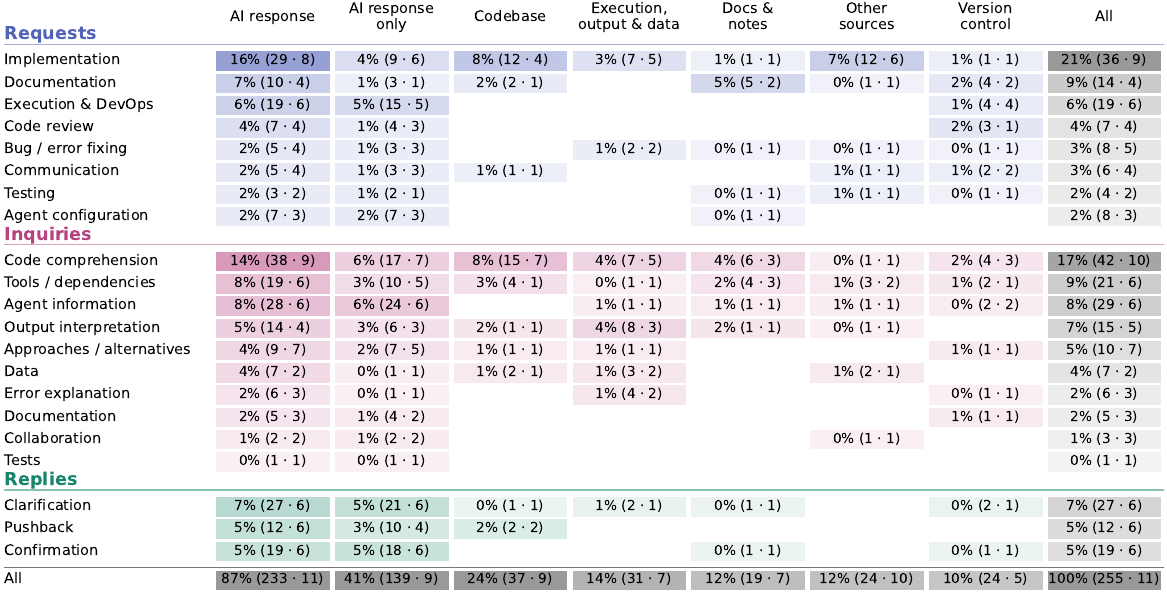}
    \vspace{-20px}
\end{table}

We classified prompts into three intents: \textcolor{requestblue}{Requests}, \textcolor{inquiryred}{Inquiries}, and \textcolor{replygreen}{Replies}.  \textcolor{requestblue}{Requests} ask the agent to produce or modify an artifact (\eg \textit{``implement the export feature''}). \textcolor{inquiryred}{Inquiries} ask the agent to retrieve or explain information, even when doing so requires tool use (\eg \textit{``why does this fail?''}). \textcolor{replygreen}{Replies} respond to an agent's question or proposal, for example by supplying requested details, confirming a choice, clarifying instructions, or rejecting a proposed approach. The key distinction between a reply and a new request or inquiry is whether the prompt continues the current goal or introduces an independent one.

Table~\ref{tab:before-to-prompt} summarizes observable activities preceding prompts, organized by prompt type. 
Rows represent prompt purposes,
grouped into \textcolor{requestblue}{Requests}, \textcolor{inquiryred}{Inquiries}, and \textcolor{replygreen}{Replies}. Columns represent
the sources developers inspected before formulating prompts,
including AI responses, code, execution output and data,
documentation and notes, other sources, and version control.
The row categories derive from our prompt coding, while the
columns group activities identified through the backward
analysis described in our analysis procedure.

For each cell, we calculate the percentage of each participant's
retained prompts that match both the row's prompt purpose and
the column's preceding activity. We then average these
percentages across all eleven participants, including zeros.
This gives each participant equal weight despite differences
in prompting frequency: P5 had five retained prompts, whereas
P6 had 43, so each qualifying prompt contributes 20\% to P5's
percentage but approximately 2.3\% to P6's. Cells show the mean percentage, followed by the prompt count
and participant count. For example, 16\% (29 $\cdot$ 8) in the
\emph{Implementation} row and \emph{AI response} column
represents 29 implementation requests across eight participants
preceded by reviewing an AI response. 
These accounted for
16\% of participants' sent prompts on average.

The \emph{AI response only} column identifies the subset
for which response review was the sole coded preceding
activity. The gray \emph{All} column summarizes prompt purposes
regardless of preceding activity, while the gray \emph{All}
row summarizes preceding activities regardless of prompt
purpose. Categories can overlap, so percentages need not
sum to 100\%. Blank cells indicate no observations, whereas
0\% indicates a positive mean percentage rounded to zero.
These activities describe what developers observably inspected
before prompting; they do not establish what information
developers explicitly provided to the agent.

\textbf{Finding 7 (\faIcon{eye}): Developers used prompts both to advance development tasks and to understand and manage the agents performing them.}
They directed agents to implement features
(21\% of participants' sent prompts on average), write
documentation (9\%), and execute code (6\%).
They also sought explanations of code (17\%), external tools and dependencies (9\%), and program output (7\%). Beyond these development tasks, developers used prompts to
ask about agents' capabilities and status, configure their
behavior, and coordinate their work.
Agent-information inquiries accounted for 8\% of prompts
on average (29 prompts across six participants), while eight prompts requested changes to agent configuration.
Developers asked whether agents could access particular
content (P4), whether they were still running
(P6, P7, P8, P11), and how many subagents were active (P8).
They also requested updates to \smalltt{SKILLS.md} (P6, P8) and instructed agents to communicate with other agents to coordinate their work (P7, P8).
Thus, developers use prompts not only to conduct regular software development, but also to manage the agents.

\textbf{Finding 8 (\faIcon{eye}): Agent dialogue frequently provided the immediate observable context for further prompting.} 
Reviewing an AI response was the \textit{\textbf{sole}} coded preceding activity for 41\% of participants' prompts on average, with no observable inspection of code, documentation, or other artifacts before the next prompt.
For example, P6 configured agents to ask clarifying questions before beginning implementation. P6 then supplied the requested details through dialogue, without observable artifact inspection between these exchanges.
Despite agents' ability to fetch the information, file, and execute the program, developers still copied and pasted the relevant information to the prompt (26\%). For 72 prompts that have explicit context, the most common were files (17), previous agent responses (13), Git (12), and terminal output (10).

\subsection{RQ2-2: Information from AI to Developers\label{sub:InfoFromAI}}
\begin{table}[t]
    \centering
    \caption{Observable practices following prompts, by prompt type. Cells show percentages of each participant's total prompts, averaged across participants, followed by (prompt count $\cdot$ number of participants).\label{tab:checks}}
    \includegraphics[width=1\linewidth]{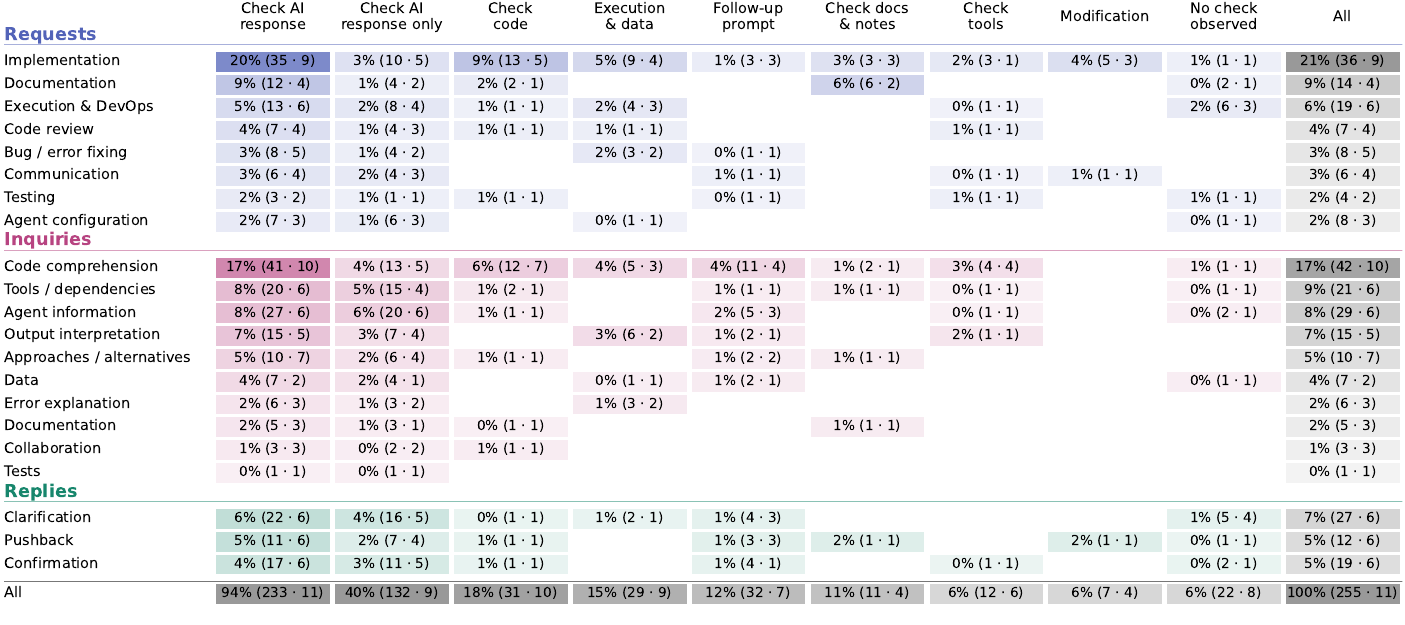}
    \vspace{-.3in}
\end{table}
Through the \textit{forward analysis} described in
\S\ref{subsubsec:analysis}, we identified observable checks following each prompt.
Table~\ref{tab:checks} summarizes these practices using the same percentage calculation and color scheme as
Table~\ref{tab:before-to-prompt}.
Practices can overlap; \emph{Check AI response only} shows prompts that response review was the sole coded subsequent practice.
We describe these actions as checks because observing them does not establish that developers validated the correctness of an agent's output. 

\textbf{Finding 9 (\faIcon{eye}): Developers varied in how they
checked agents' outputs, sometimes asking agents to reassess
their own responses or work.}
Checking AI responses was the \textbf{\textit{sole}} coded
subsequent practice for \textbf{40\%} of participants' retained
prompts on average, ranging from 0\% (P1, P5) to 72\% (P7).
This practice followed more than half of retained prompts for
P3, P6--P8, and P10, but fewer than 16\% for P1, P2, P5,
and P9. Developers also opened code files (18\%), executed
programs or checked data (15\%), and checked documentation
and notes (11\%). These percentages summarize checks observed
within the analysis interval and do not rule out later checking.

In some cases, developers asked agents to check their preceding
responses or work through \emph{follow-up verification prompts},
such as \cSay{are you sure? (P11).}
These occurred after 12\% of participants' retained prompts
on average. For example, after reviewing plans proposed by
the agent, P6 instructed it to check its own logic against
a reference configuration. Similar practices occurred among
P3, P4, P6, P8, P9, P10, and P11.
These examples show developers extending the agent's role
from producing outputs to checking them by specifying what
it should examine or questioning its preceding answer.
\begin{wraptable}{r}{0.5\textwidth}
\centering
\footnotesize
\setlength{\tabcolsep}{3pt}
\renewcommand{\arraystretch}{1.05}
\caption{Prompt and task-switching characteristics by pattern.
Percentages, task counts, and switch counts are averaged across
participants; prompt counts are medians of retained prompts.
\emph{Response only} denotes prompts followed solely by
AI-response checking. Prompt-intent categories can overlap.
\label{tab:pattern-prompts}}
\begin{tabular}{@{}lrrrr@{}}
\toprule
& \makecell{\supervisor\\{\scriptsize P1, P2, P5}}
& \makecell{\delegator\\{\scriptsize P6--P8}}
& \makecell{\learner\\{\scriptsize P3, P4, P9}}
& \makecell{Mixed\\{\scriptsize P10--P11}} \\
\midrule
Mean tasks & 5.3 & 11.7 & 3.3 & 2.0 \\
Mean switches & 27.0 & 79.7 & 18.3 & 11.0 \\
\midrule
Median prompts & 11 & 42 & 21 & 16.5 \\
Response only & 5\% & 69\% & 39\% & 49\% \\
\textcolor{replygreen}{Replies}
& 7\% & 30\% & 12\% & 18\% \\
\textcolor{inquiryred}{Inquiries}
& 18\% & 39\% & 76\% & 45\% \\
\textcolor{requestblue}{Requests}
& 78\% & 35\% & 18\% & 36\% \\
Agent info/config.
& 0\% & 23\% & 9\% & 9\% \\
\bottomrule
\end{tabular}
\vspace{-20px}
\end{wraptable}

\textbf{Finding 10  (\faIcon{comment}): Developers differed in how much they checked agents' works, and they described three recurring bases for deciding how closely to check agents' work: (1) \textit{harnesses} set before the agent ran (P5, P6, P7, P9, P11), (2) their own \textit{expertise} (P1, P4, P5, P11), or (3) \textit{trust} in the model
(P2, P6, P8, P10, P11)}, when asking their verification strategies in the follow-up interviews.
Those who relied on \textit{harnesses} constrained agents' actions in advance to reduce the need for subsequent inspection.
This took several forms: writing tests that passing output had to
satisfy (P5, P11), scoping the agent's task tightly (P6, P9),
aligning on a plan or configuring the agent to ask before
proceeding (P6, P7), and requiring observable proof such as a
before-and-after recording (P7). P6 and P7 went furthest,
auto-merging well-defined changes without review---\cSay{if the
agent thinks this is safe to land, then let it land} (P7). As P9
put it, they kept the agent's \cSay{job minimal\ldots{} whatever it
does, it has a less chance of messing up.}
Those who relied on their own \textit{expertise} used it to decide which parts of the output to look at rather than reviewing all of it (P1, P4, P5, P11). P5 checked \cSay{some of the
performance or behavior-critical parts that AI wrote,} guided
by knowledge of how new edits would ripple through the system; P4 used domain knowledge to catch a redundant configuration code, and P11 acted on a \cSay{gut feeling} that something was off before reading any code. 
P1, being the senior developer, highlighted the challenges in reviewing junior's code: \cSay{the code to review is a lot but the quality is way less.}

Participants (P6, P8, P10, and P11) described adjusting their \textit{trust} in AI agents over time and correspondingly adapting how closely they checked the agents' outputs.
For P6 and P10, trust earned
over time lowered how much they checked. P6 reviewed
less as \cSay{trust is bigger} on repetitive tasks, and P10
described trust \cSay{built up over time} until evaluating outputs
was the only check they still ran. In contrast, despite orchestrating multiple agents, P8 \cSay{still don't trust the models that much}, and read every pull request, adding an adversarial AI review on top. 
Two accounts bound this relationship: P11 trusted the agent
\textit{more} in unfamiliar domains, having less basis to
override it, and P2 noted a ceiling that trust cannot resolve: \cSay{I can write some tests, like maybe ten, twenty tests myself, and then that test might pass, but it doesn't guarantee me.}

\subsection{RQ2-3: Management Patterns: \supervisors, \delegators, and \learners}
Whereas RQ1-1 and RQ1-2 describe developers' overall time
allocations and task switching, these aggregate findings do
not show how developers combine providing information to
agents with

\clearpage
\begin{wrapfigure}{l}{0.442\columnwidth}
\centering
    \includegraphics[width=\linewidth]{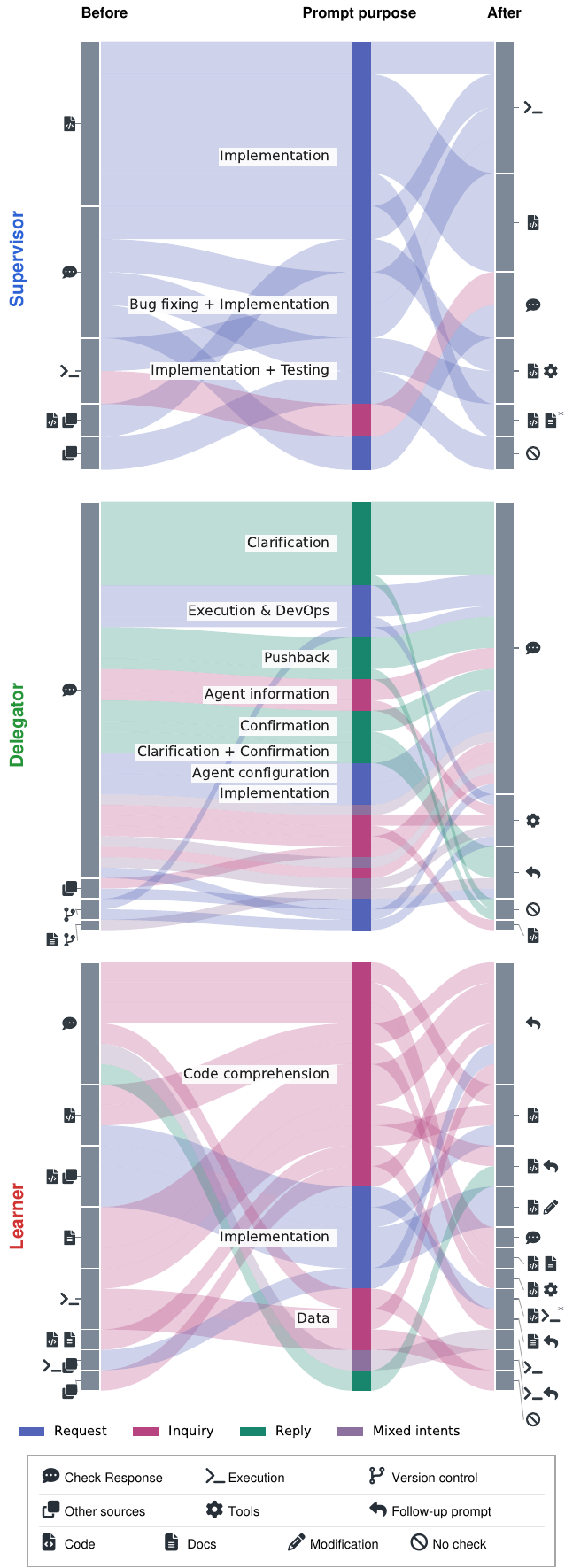}
    \caption{Management patterns for P2 (\supervisor), P6 (\delegator), and P9 (\learner) from top to bottom. Colors denote \textcolor{requestblue}{Requests}, \textcolor{inquiryred}{Inquiries}, \textcolor{replygreen}{Replies}, and mixed intents (purple).  Adjacent icons indicate co-occurring activities.}
    \label{fig:management-patterns}
\end{wrapfigure}
\noindent 
examining their outputs. 
We therefore connect preceding activities, prompt types, and subsequent activities to identify patterns of working with agents. 
After constructing participant-level prompt-centered flows, the researchers compared participants along three dimensions evident in the observational data: (1) dominant prompt purposes (request, inquiry, and reply); (2) the extent to which checks extended beyond reading the agent response, measured by the ratio of response-only checks to checks involving other artifacts or execution; and (3) the degree of task interleaving and agent-oriented coordination (Figure~\ref{fig:interleaved}).
Through iterative comparison across these dimensions, we identified three recurring patterns.
We then examined participant-level distributions to assess whether the coded data supported these qualitative distinctions.
Sessions that combined characteristics of multiple patterns were retained as mixed cases.
These patterns characterize the observed sessions and are not intended to define exhaustive or mutually exclusive developer types.
The criteria used to identify the patterns are provided in the replication package.

\textbf{Finding 11: Our participants exhibited three recurring patterns
of managing AI agents: \supervisors, \delegators, and \learners.}
Figure~\ref{fig:management-patterns} presents Sankey diagrams for
P2, P6, and P9, illustrating the \supervisor,
\delegator, and \learner patterns, respectively.
For each participant, the left
column shows activities preceding
a prompt, the center column shows prompt purposes (\textcolor{requestblue}{Requests}, \textcolor{inquiryred}{Inquiries}, and
\textcolor{replygreen}{Replies}), and the right column shows
activities following the prompt. 
Flow widths represent the
proportion of that participant's prompt-centered episodes following
each path.
When AI-response review co-occurred with another activity,
we grouped the episode by the others; AI-response review is shown only when it was the sole coded activity.

\supervisors (P1, P2, P5) primarily used agents to produce or
modify artifacts while remaining directly involved in checking the
result. P2 illustrates this pattern in
Figure~\ref{fig:management-patterns}: implementation requests were
followed by activities such as inspecting code, examining generated
changes, or executing the resulting artifact. Across P1, P2, and P5, requests accounted for an average of 78\%
of prompts, compared with 18\% inquiries and 7\% replies.
Only 5\% of their prompts were followed \textbf{solely} by checking the AI response, indicating that their checking
usually extended beyond reading the agent's textual response to
inspecting or exercising the resulting artifact. These participants
worked on an average of 5.3 tasks and made 27.0 task switches per
session. 

Thus, although \supervisors also interleaved work across tasks,
their management remained centered on directly inspecting what the agent produced.

\delegators (P6--P8) showed a contrasting pattern in which
management occurred primarily through interaction with agents rather
than direct inspection of their artifacts. P6's session in
Figure~\ref{fig:management-patterns}, for example, involved repeatedly
moving among agent sessions, responding to clarification questions,
and directing ongoing work, with little code inspection or execution
following prompts. \delegators worked on substantially more distinct tasks than the
other groups, averaging 11.7 tasks and 79.7 task switches per session. Their prompts were also more conversational:
39\% were inquiries, 35\% requests, and 30\% replies. In addition,
23\% of their prompts concerned agent information or configuration.
Most notably, 69\% of their prompts were followed \textbf{solely} by checking
the AI response. Together, these characteristics show a management
pattern centered on dialogue and coordination: \delegators supplied
clarifications, monitored agents through responses, configured
agent behavior, and delegated more of both implementation and checking
to the agents themselves.

\learners (P3, P4, P9) primarily used agents to understand unfamiliar artifacts, tools, and even data. 
P9 illustrates this
pattern in Figure~\ref{fig:management-patterns}, where most of the prompts asked the agent to explain the repository and its surrounding context rather than to produce a new artifact. 
Consistent with this pattern, inquiries accounted for 76\% of prompts among \learners, compared with 18\% requests and 12\% replies. 
39\% percent of their prompts
were followed solely by checking the AI response. 
In these cases, the response itself was often the information being sought rather than an
intermediate artifact requiring further inspection. \learners worked on an average of 3.3 tasks and made 18.3 task switches per session, showing less task interleaving than \delegators. 
Their interaction with agents was primarily centered on building understanding rather than directing implementation.

The patterns could also overlap within a session. P9, for example, primarily behaved as a \learner while exploring an unfamiliar repository, but later shifted toward a \supervisor pattern in another task when
requesting changes and inspecting the resulting artifacts. 
P10 and P11 similarly combined \supervisor and \learner behaviors while working with external libraries and applying them to their own projects. 
Together, they worked on an average of 2.0 tasks and made 11.0 task switches per session; their prompts included both inquiries (45\%) and requests
(36\%), while 49\% were followed solely by AI-response checking.
We therefore interpret \supervisor, \delegator, and \learner as recurring patterns of agent management that can shift with the task and context, rather than as fixed developer types.

\subsection{RQ2-4: Information from AI vs. Colleagues}

Developers spent an average of 9.9\% of their observed session time on messaging and email.
Although some communication content was visible in the recordings, we did not analyze it to protect participants' and their colleagues' privacy.
We therefore report communication time without characterizing the information exchanged. 
Instead, the follow-up interviews asked what information developers preferred to seek from AI agents versus colleagues.

\textbf{Finding 12 (\faIcon{comment}): Developers turned to colleagues when questions required organizational context, human authority, or collective alignment, while using agents for technical information accessible from code, documentation, or public knowledge.}
Participants' accounts highlighted two distinctions.
The first concerned \textit{organizational context and human authority}.
Developers consulted colleagues about the intent behind feature requests (P4, P8), historical reasoning not apparent in code (P1, P7), permissions requiring human authority (P8), and the meaning of domain-specific concepts in their work (P6).
P8 summarized this distinction: \cSay{I mostly only talk to humans now about intent and access.}
P1 emphasized colleagues' knowledge of
\cSay{historical reasons why certain things had to be done.}
In contrast, developers described turning to agents for technical
questions answerable from code, documentation, or public knowledge
(P1--P4, P7, P8, P9, P11).
P4 distinguished requirements from implementation guidance:
\cSay{requirements come from a colleague...but with AI, it's more
like how can I accomplish this?}
P7 described their team's expectation:
\cSay{if a agent can reasonably
answer it, then you should ask an agent.} The second distinction concerned \textit{collective decision}. Developers consulted colleagues about scope, system
design, and coordination over who was working on what (P3, P4, P7--P9, P11). P11 explained:
\cSay{if this question requires more like alignment, then I'll reach out to my colleague first
before I reach out to AI.}

Developers spent an average of 7.3\% of their observed session time
reviewing or editing documentation. This contrasts with the pre-AI survey
by Stettina \& Heijstek~\cite{10.1145/2038476.2038509}, in which half of
respondents reported spending less than 15 minutes per day maintaining
documentation despite recognizing its importance.

\textbf{Finding 13 (\faIcon{comment}): Documentation served dual purposes in agent-assisted
development: preserving knowledge for people and providing context for
agents.} 
During interview, P6 questioned the premise directly: if AI could explain codebase from code alone, \cSay{What's the point of
documenting more?} finding it \cSay{absolutely absurd that one person writes a lengthy documentation that then the next person feeds into the AI to shorten it into a summary.}
P7 offered two reasons that covered most participants: \cSay{docs maybe serve two purposes
pre-AI, (1) help you understand things at a
high level but then also (2) very mechanically like how to implement something ..., that [(2)] just for agents at this point.} Both reasons summarize across participants.

The first was preserving knowledge that might otherwise be lost or remain implicit (P1, P3, P5). P1 and P5 adopted structured pre-work documents to maintain shared understanding as AI accelerated code production, while P3 configured an agent to log design and architectural decisions in real time. As P5 described, this involved \cSay{distilling yourself and extracting your information into a markdown document so that everyone else can take up on the task.} The second was providing context for agents. P2 moved team documentation into a repository so Claude could access it and check its consistency with the code. AI also lowered the effort of maintaining documentation (P2, P4, P5, P7, P8, P10, P11): participants used agents to generate wiki pages, notes, and drafts, though outputs still required cleanup. As P8 put it, \cSay{the agent writes all the docs up front}, but they still had to \cSay{deslop the docs all the time}.

\section{Discussion}
\subsection{Developers as Managers of AI Agents}
Our findings show not only \emph{how much time} developers spend
working with underlying artifacts and interacting with agents
, but also \emph{where they exercise attention and control}
when working with coding agents (Finding~1--6).
The \supervisor, \delegator, and
\learner patterns (Finding~11) make this redistribution visible:
\supervisors remained close to generated artifacts, \delegators managed
work primarily through dialogue and coordination, and \learners used
agents to build understanding.

Across these patterns, agent management involved three recurring demands:
\textit{bounding agent autonomy}, \textit{exercising technical judgment}, and allocating
attention across concurrent delegated work. Kalliamvakou~\etal's account
of effective engineering management provides a useful lens for bounded
autonomy and technical judgment \cite{8094304}, while prior work on coding
agents has documented developers' continued use of planning, supervision,
and validation \cite{huang2025professional}. Our observations extend these
accounts by showing that control can occur at different points in the
workflow.

First, developers bounded agent autonomy through task scope, tests,
careful plans, and checked outputs
through response review, artifact inspection, execution, and follow-up
prompts (Findings~9--10). Control therefore occurred not only after
generation, but also upstream by harnessing the conditions under which agents operated.

Second, developers relied on technical judgment, or \cSay{gut feeling (P11)} to decide when closer
inspection or additional investigation was warranted (Finding~10).
The \learner pattern further shows that agents could help developers
build the understanding needed to make those judgments.

Third, coding agents introduced additional parallelism into already
fragmented software work \cite{10.1145/3210459.3210471,mark2023attention,10.1145/2635868.2635892}.
Participants used deliberate task switching, reminders, and external
notes to manage concurrent agent-mediated work (Findings~5--6). Among
\delegators, control was exercised less through artifact
inspection and more through deciding when human attention should return.

Agent management also remained embedded in workplace collaboration.
Prior work has shown that developers' information needs involve both
technical artifacts and coordination with others
\cite{ko_information_2007,begel2009coordination}.
Our participants used agents mainly for technical information, while
turning to colleagues for organizational history, intent, authority,
and collective alignment (Finding~12). Agent capability may therefore
shift, rather than eliminate, human information seeking. Taken together, our findings characterize agent management as a
redistribution of human control: developers constrain agents, decide when
intervention is needed, and direct limited attention across concurrent work.

\subsection{Designing for Supervising, Coordinating, and Learning with AI Agents}

Our findings suggest that development tools should support not only the
execution of programming tasks but also the work involved in managing AI
agents. 
The \supervisor, \delegator, and \learner patterns reveal
different needs: visibility and verification support for generated
artifacts~\cite{yang2026whensoftwarefactorylightsoff, bacchelli2013expectations},
coordination across concurrent agent work~\cite{10.1145/3706598.3713581, begel2009coordination}, and support for building understanding~\cite{10.1145/3597503.3639187}.

For \supervisors, management depended on inspecting generated artifacts, executing programs, and deciding which changes warranted closer review.
As P1 remarked, \cSay{Sometimes, 80\% of my day is code review}.
These practices suggest opportunities for tools that make verification evidence more visible and accessible by linking agent-generated changes to relevant artifacts, highlighting complex or fragile parts of the software that may warrant closer inspection~\cite{yang2026whensoftwarefactorylightsoff}, and presenting executed tests and their results~\cite{10.1145/3765754}.
Tools could also help developers locate changes that implement specific requirements or business rules and connect them to the intended behavior and relevant tests, supporting review of whether the generated logic matches domain expectations.
Beyond presenting evidence, tools could challenge developers to examine their assumptions~\cite{sarkar_ai_2024}, identify remaining uncertainties, and formulate questions about the generated work, since effective questioning depends partly on recognizing what is not yet understood~\cite{miyake_ask_1979}.
These interfaces could help developers assess an agent's report of completion and decide where to direct their review effort.

For \delegators, the central challenge was less inspecting each artifact
than tracking multiple ongoing agent tasks and deciding when human
attention was needed. Participants improvised with reminders, notes,
and deliberate switching strategies (Findings~5--6), suggesting
opportunities for coordination interfaces that indicate which agent
tasks are progressing independently, blocked, awaiting approval, or
otherwise require human intervention. Such interfaces could build on
prior work on workspace awareness, dependencies, and traceability
in software development~\cite{sarma2003palantir, 4639062, 5070505}.
Furthermore, some developers explicitly directed agents to exchange
information or coordinate their work (Finding~7), suggesting
opportunities to expose dependencies and information flow across
concurrent agent sessions~\cite{5070505}.
Future studies should examine whether frequent switching among agent
tasks affects cognitive load, task resumption, and developers'
perceived control~\cite{10.1145/3847122}, and evaluate whether
coordination interfaces can mitigate the \cSay{context switching hell} (P2)~\cite{card1986multiple}.

For \learners, agents became part of developers' process of
understanding unfamiliar codebases, tools, and data, extending prior
observations of LLM-assisted code comprehension by Nam~\etal~
\cite{10.1145/3597503.3639187}. Tools could therefore connect
agent-generated explanations to underlying artifacts, helping developers inspect and contextualize the information they receive.
Furthermore, participants' accounts of documentation serving both human and agent needs (Finding~13) suggest opportunities to preserve the evidence, reasoning, and actively elicit tacit knowledge behind software
artifacts from developers~\cite{10.1145/3377811.3380405, ryan2013acquiring}.
Building on our previous work~\cite{10.1145/3797105}, future tools
could also provide personalized pedagogical support that draws on
the developer's task context and helps establish shared
understanding between developers and agents~\cite{schelble2022let}.

Rather than assuming a single interaction model for agentic
development, these patterns suggest that tools should support
different forms of human involvement as developers move between
supervising artifacts, coordinating delegated work, and building
understanding. Because these patterns can overlap and shift within
sessions, such support should adapt to the task and context rather
than treating developers as fixed types.

\section{Conclusion}

Through workplace shadowing and follow-up interviews with 11 professional developers, we examined how developers allocate attention, direct coding agents, and check their outputs during everyday work. Developers spent an average of 33.4\% of observed time formulating prompts or checking AI responses. Within approximately 90-minute sessions, they worked on 1–17 distinct tasks, with switching rates ranging from 0 to 101.6 switches per hour. We identified three recurring patterns, \supervisor, \delegator, and \learner, that combined prompt purposes, task interleaving, and checking practices in different ways: directly inspecting generated artifacts, coordinating delegated work through chats, and seeking explanations to build understanding. Participants also described using agents for technical information while turning to colleagues for organizational context, authority, and collective alignment. Future work should deepen our understanding of how developers manage attention across concurrent agent tasks, and inform tool designs that support verification, coordination, and learning.

\section{Data Availability}
Our replication package~\cite{replicationpackage} contains the
codebooks and coding guidelines used for the observational activity
coding and prompt-centered analysis. We also release the anonymized
sequences of coded activities for all participants, which support the
reported analyses of time allocation, task switching, and activities
surrounding prompts. In addition, the package provides participant-level
visualizations corresponding to the figures reported in this paper,
including the complete-session task visualizations and management-pattern
visualizations for all 11 participants.
\bibliographystyle{ACM-Reference-Format}
\bibliography{references,sample-base,ref}

\end{document}